\magnification=1200
\tolerance=10000
\centerline{{\bf INFRARED SINGULARITIES IN THE}}
\centerline{{\bf RENORMALIZATION GROUP FLOW OF}}
\centerline{{\bf YANG-MILLS THEORIES IN THE AXIAL GAUGE}}
\bigskip
\centerline{
{\bf Annamaria Panza}
and 
{\bf Roberto Soldati}}
\centerline{{\sl Dipartimento di Fisica "A. Righi", Universit\`a di Bologna,}}
\centerline{{\sl Istituto Nazionale di Fisica Nucleare, Sezione di Bologna,}}
\centerline{{\sl 40126 Bologna - Italy}}
\vskip 2.0 truecm
\centerline{Abstract}
\bigskip
\noindent
{\it It is shown, by explicit calculation in the axial gauge, that the renormalization group flow
for the Wilson loop in perturbation theory does exhibit singularities and consequently it can not eventually 
reproduce the gauge invariant result,  when the
infrared cut off is removed.}
\vskip 2.0 truecm
\noindent 
{\bf 1.}\quad 
Since the appearance of Wilson's seminal ideas [1] and subsequent Polchinski's developments [2],
the Renormalization Group Flow (RGF) approach to the quantization and renormalization of field 
theories has become more and more popular, both from perturbative and non-perturbative points 
of view. A truly remarkable amount of work has been done in the last few years along that line
[3] and, in particular, the RGF approach to the setting up of perturbative Yang-Mills theories
in covariant gauges has been thoroughly discussed in [4]. There, it is neatly pointed out that,
in the presence of some continuous local symmetry such as the non-abelian gauge symmetry, 
the main problem is the solution of the so called
fine tuning functional equation, which corresponds to the restoration of the symmetry at the quantum level, 
after removal of the ultraviolet and infrared regulators $\Lambda_0$ and $\Lambda$ respectively.
The solution of the above functional equation turns out to be either rather burdensome 
at the perturbative level (technically extremely heavy beyond one loop)
or insofar unknown at the non-perturbative level.
Later on it has been argued [5,6] that the above mentioned difficulty could be circumvented after 
transition to the axial gauge. As a matter of fact, it turns out that
\quad {\it (i)} regularized Green's
function in the axial gauge fulfill, in the limit $\Lambda\downarrow 0$,
Lee-Ward-Takahashi identities much simpler than Slavnov-Taylor
identities owing to decoupling of Faddeev-Popov ghosts;\quad {\it (ii)} the presence of the infrared cut off does
provide a natural way to screen the non-covariant singularities of the Green's functions and, 
specifically, of the free propagator.
Quite remarkably, it has been recently
proved [6] that a modified but simple form of the Lee-Ward-Takahashi identities in the axial gauge
can be maintained                   
to all scales $\Lambda\not=0$, order by order in perturbation theory, leading thereby to 
a mild breaking of the non-abelian 
gauge symmetry of the RGF. In order to obtain that result, the infrared cut off $\Lambda$
was suitably identified with a vector boson mass. Actually, according to universality
of the renormalized theory in the limit $\Lambda\downarrow 0$, any form of the intermediate
infrared regulator  must eventually lead, once removed, to the very same renormalized theory in which the 
gauge symmetry has to be fully restored. Now, it turns out that, at least in the framework
of perturbation theory, some gauge fixing Action involving a mass term of the vector boson really 
represents the simplest tool
to investigate the infrared structure of the regulated theory and to check, order by order,
the consistency of the RGF procedure, what has been done [5,6] concerning the one loop $\beta$-function.  

In spite of the welcome benefit to have simple
modified Lee-Ward-Takahashi identities to any order in perturbation theory,  the axial gauge RGF approach has been 
suspected to exhibit singularities [7] in the physical limit $\Lambda\downarrow 0$. 
In this note we actually prove with an explicit perturbative calculation that the usual expression of the euclidean Wilson 
loop in the infrared cut off formulation does not flow smoothly to the gauge invariant result, after removal 
of the infrared cut off itself.
More specifically, we shall see that the Wilson loop, up to the fourth order in the non-abelian coupling,
does not admit a smooth limit when $\Lambda\downarrow 0$, {\it i.e.}, infrared singularities do not cancel even 
for a quantity that is formally gauge invariant at the physical limit. 
\bigskip
\noindent 
{\bf 2.}\quad 
In the spirit of the RGF formulation, the euclidean non-abelian gauge theory
can be generally defined in perturbation theory by the usual gauge invariant Action - 
to our purposes we can neglect matter fields -
supplemented by some suitable massive term involving the necessary infrared cut off, 
together with some gauge fixing Action which allows to
study the physical limit of vanishing infrared cut off. In the present investigation 
we shall consider the axial gauge fixing which is specified,
in the euclidean formulation, by a fixed vector $n_\mu$, in such a way that our starting 
euclidean Action becomes [6]       
$$
{\cal A}[A_\mu]=\int d^4 x\left\{ {1\over 2}{\tt tr}\left[F_{\mu\nu}(x)F_{\mu\nu}(x)\right]+
\Lambda^2{\tt tr}\left[A_\mu(x) A_\mu(x)\right] + n_\mu{\tt tr}\left[A_\mu(x)\lambda(x)\right]\right\}\ ,
\eqno(1)
$$
where trace is over products of $SU(N)$ Lie algebra valued euclidean vector potentials, field strengths and the auxiliary 
field which enforces the axial gauge choice $n_\mu A_\mu=0$. Notice that the Faddeev-Popov ghosts can here be disregarded,
even in the non-abelian case, thanks to their decoupling within the axial gauge choice.
The momentum space euclidean free vector propagator is readily obtained to be ($n^2\equiv n_\mu n_\mu$)
$$
\eqalign{
&\tilde D_{\mu\nu}(k;\Lambda,n)={1\over k^2+\Lambda^2}\times\cr
&\left\{\delta_{\mu\nu}-(n\cdot k){n_\mu k_\nu+n_\nu k_\mu\over
(n\cdot k)^2+n^2\Lambda^2}+n^2{k_\mu k_\nu\over (n\cdot k)^2+n^2\Lambda^2}
-\Lambda^2{n_\mu n_\nu\over (n\cdot k)^2+n^2\Lambda^2}\right\}\ ,\cr}
\eqno(2)
$$
which manifestly enjoys 
$$
n_\mu\tilde D_{\mu\nu}(k;\Lambda,n)=0\ .
\eqno(3)
$$ 
It can be easily checked that, owing to the presence of the infrared cut off $\Lambda$, the above euclidean propagator
corresponds to the presence of two independent polarizations and is apparently free from small momenta singularities.
Furthermore, since the three and four point elementary non-abelian vertices are the usual ones and it can be shown that
the propagator (2) does fulfill {\it na$\ddot\imath$ve} power counting [8], it turns out that the present model 
is power counting renormalizable.
Of course, this can be obtained at the price of breaking the euclidean $O(4)$ symmetry and, moreover, it can be explicitely
verified that the above propagator (2) does not admit a well defined limit when $\Lambda\downarrow 0$, even in the weak topology
of the tempered distributions. This means that, on the one hand, the introduction of the Wilson infrared cut off together with
the axial gauge subsidiary condition allow for the setting up in perturbation thery of a well defined set 
of renormalized Schwinger's function 
which, however, do break $O(4)$ symmetry and do not admit a smooth limit at the
physical gauge invariant point $\Lambda=0$, even within the weak topology of the tempered distributions. 
Nonetheless, one can rather reasonably
expect that, at least for some suitable renormalized quantities which formally become gauge invariant at 
the physical point $\Lambda=0$,
the corresponding renormalization group flow is smooth in the limit of vanishing infrared cut off and 
does reproduce the gauge invariant result, up to
renormalization prescriptions. In the non-abelian case such a kind of behaviour could be exhibited by 
geometrical path-ordered phase factors, because
$S$-matrix elements among fundamental fields are affected by severe infrared divergences 
(generally unmanageable in perturbation theory).

In the present paper we shall analyze the perturbative expansion of the euclidean average path-ordered phase factor
$$
\eqalign{
W_\Gamma(\Lambda)&={1\over N}\left<{\tt tr}\left\{P\exp\left[ig\oint_\Gamma dx_\nu A_\nu(x)\right]\right\}\right>\cr
&\equiv{\cal N}^{-1}\int [dA_\nu]\exp\{-{\cal A}[A_\nu]\}{1\over N}{\tt tr}\left\{P\exp\left[ig\oint_\Gamma dx_\nu A_\nu(x)\right]\right\}
\ ,\cr}
\eqno(4)
$$
where $\Gamma$ is a closed
rectangular path lying on the $Ox_3x_4$ plane, centered at the origin and with sides of lengths $2L$ and $2T$
along the $Ox_3$ and $Ox_4$ axes respectively, whilst we have set ${\cal N}\equiv\int [dA_\nu]\exp\{-{\cal A}[A_\nu]\}$.
In eq.~(4) the vector potential is supposed to belong to a fundamental representation of the $SU(N)$ Lie algebra, {\it i.e.} we set
$A_\mu(x)=A_\mu^A(x)T_F^A$. 
The formal quantity in eq.~(4) will be referred to as the infrared cut off Wilson loop in the axial gauge.
The perturbative expansion of the quantity (4) is well defined in terms of the propagator (2) and of the usual euclidean Feynman's
rules for the three and four point vertices of the pure Yang-Mills theory. This is true provided some regularization is 
introduced in order to deal with ultraviolet infinities. In this note we shall adopt dimensional regularization and the 
dimensionally regularized quantity corresponding to eq.~(4), order by order in perturbation theory, 
will be denoted by ${\tt reg}W_\Gamma(\Lambda)$.
This regularized quantity crucially depends upon the infrared cut off $\Lambda$, at least in perturbation theory.
In fact, as already mentioned, the free propagator (2) does not admit a well defined limit $\Lambda\downarrow 0$ in the weak topology of
the tempered distributions. This means that in general many individual graphs, which contribute to the
quantity ${\tt reg}W_\Gamma(\Lambda)$ at a given order, will diverge in the limit $\Lambda\downarrow 0$. 
Nonetheless, one might conjecture that the full quantity ${\tt reg}W_\Gamma(\Lambda)$ keeps a finite value in the physical limit
$\Lambda\downarrow 0$, order by order in perturbation theory. Furthermore, if this were true, it might also happen
that the possible finite value of the quantity ${\tt reg}W_\Gamma(0)$ 
could also be independent from the specific choice 
of the axial gauge fixing. In this case, the gauge invariant value of the dimensionally regularized euclidean Wilson loop
could eventually be obtained, starting from the perturbative RGF approach in the axial gauge.
The above conjecture can be actually verified at the lowest order ${\cal O}(g^2)$, as we shall discuss in the next section. 
It is clear that to check the above conjecture at higher orders does represent a highly non-trivial test. In this note we shall show that the
basic quantity ${\tt reg}W_\Gamma(\Lambda)$ does not admit a limit when $\Lambda\downarrow 0$ already at the fourth order in $g$.
\bigskip
\noindent 
{\bf 3.}\quad   
Let us first check what happens at the lowest order in perturbation theory. As already mentioned, we shall regulate 
the ultraviolet divergences by means of dimensional regularization, 
in such a way that the regularized Green's functions actually fulfill
the modified Lee-Ward-Takahashi identities [7]. To start with, we recall that
in the case of euclidean massless $SU(N)$ gauge theories, the covariant gauge
result ${\cal O}(g^2)$ in $2\omega$-dimension reads
$$
\eqalign{
&{\tt reg}W_2(L,T) =
-g^2{\cal C}(\omega)
\left\{I_\omega(L)+I_\omega(T)-J_\omega(L,T)-J_\omega(T,L)\right\}\ ,\cr
&{\cal C}(\omega)\equiv 4C_2(F) \mu^{4-2\omega}(2\pi)^{-2\omega}\ ,\cr}
\eqno(5) 
$$
where $C_2(F)$ is the quadratic Casimir operator of the fundamental representations of $SU(N)$
whilst 
$$
\eqalignno{
I_\omega(L)&=\int d^{2\omega-1}{\bf k}\int_{-\infty}^{+\infty}
{\sin^2(k_3L)\over k_3^2}{dk_3\over k_3^2+{\bf k}^2}\ ,&(6a)\cr
I_\omega(T)&=\int d^{2\omega-1}{\bf k}\int_{-\infty}^{+\infty} 
{\sin^2(k_4T)\over k_4^2}{dk_4\over k_4^2+{\bf k}^2}\ ,&(6b)\cr
J_\omega(L,T)&=\int_{-\infty}^{+\infty} dk_3\int_{-\infty}^{+\infty} dk_4
\int d^{2\omega-2}k_\perp\ {\sin^2(k_3L)\over k_3^2}
{\cos(2k_4T)\over k_\perp^2+k_3^2+k_4^2}\ .&(6c)\cr}
$$
After setting $\sigma\equiv 4LT$ - the area of the rectangular contour -
and $\beta\equiv (L/T)$, we find 
$$
\eqalignno{
&I_\omega(L)=\pi^\omega L^{4-2\omega}{\Gamma(\omega -2)\over 2\omega-3}\
,&(7a)\cr
&J_\omega(L,T)=\pi^\omega T^{4-2\omega}\sum_{n=1}^\infty
(-)^{n+1}{\Gamma(n+\omega-2)\over (2n-1)n!}\beta^{2n}\ ,\qquad \beta^2\le 1\
.&(7b)\cr}
$$
Notice that, in the limit $\omega\uparrow 2$ and  after analytic
continuation, one eventually finds for arbitrary $\beta^2>0$
$$
\lim_{\omega\uparrow 2}J_\omega(L,T)\equiv
J_2(\beta)=2\pi^2\left(\beta\arctan\beta-\ln\sqrt{1+\beta^2}\right)\ ,
\eqno(8) 
$$
where we understand the inverse trigonometric functions to be their principal branches.
It follows that, in the massless case, the ${\cal
O}(g^2)$ result around $\omega=2$ can be written as
$$
\eqalignno{
&{\tt reg}W_2(L,T)\buildrel \omega\uparrow 2 \over \sim
W_2^{\rm div}+W_2^{\rm fin}(L,T)\ ,\cr
&W_2^{\rm div}=2\hat g^2C_2(F){1\over \epsilon}\ ,\qquad
\epsilon\equiv 2-\omega\ ,\quad \hat g\equiv {g\over 2\pi}\ ,&(9a)\cr 
&W_2^{\rm
fin}(L,T)= 2\hat g^2C_2(F)\times\cr
&\left\{\gamma_E+1+\ln\pi+2\ln\left({\sigma\mu\over d}\right)
+\beta\arctan\beta+{1\over \beta}{\rm arccot}\beta
\right\}\ ;&(9b)\cr
&d\equiv 2\sqrt{L^2+T^2}\ .\cr}
$$
From the above expressions it follows that, if we add the ${\cal O}(g^2)$
counterterm
$$
W_2^{\rm c.t.}=-2\hat g^2C_2(F)\left({1\over
\epsilon}+\gamma_E+1+\ln\pi\right)\ , \eqno(10)
$$
we eventually obtain the renormalized euclidean Wilson loop in its minimal
form, up to the order $g^2$: namely,
$$
\eqalign{
W_2^{\rm ren}(\beta,\sigma)&=2\hat
g^2C_2(F)\left\{2\ln\left({\sigma\mu\over d}\right) +
\beta\arctan\beta+{1\over \beta}{\rm arccot}\beta
%\sum_{n=1}^\infty
%{(-)^{n+1}\over 2n(2n-1)}\left(\beta^{2n}+\beta^{-2n}\right)
\right\}\cr
&\buildrel \beta\ll 1 \over \sim
C_2(F){g^2T\over 4\pi L}=-TV_2(L),\quad V_2(L)=-g^2{C_2(F)\over 4\pi L}\ .\cr}
\eqno(11)
$$
Notice that the quantity $V_2(L)$ just corresponds to the lowest order Coulomb potential
of the non-abelian gauge theory [9].

Let us now turn to the same calculation in the massive case using the euclidean axial gauge
$n_\mu A_\mu=0$. From the general expression (2) and 
after choosing, {\it e.g.}, $n_1=n_2=n_3=0,\ n_4=1$ we find that the only
non-vanishing and relevant component of the infrared cut off vector propagator is
$$
\tilde D_{33}(k;\Lambda,n)={1\over k^2+\Lambda^2}\left\{1+{k_3^2\over
k_4^2+\Lambda^2}\right\}\ .
\eqno(12)
$$
As a consequence, we can write the dimensionally regularized and infrared cut off ${\cal
O}(g^2)$ Wilson loop in the axial gauge according to
$$
{\tt reg}\hat W_2(\Lambda;L,T) =
-g^2{\cal C}(\omega)
\left\{I_\omega(\Lambda;L)+\hat I_\omega(\Lambda,T)-
J_\omega(\Lambda;L,T)-\hat J_\omega(\Lambda;T,L)\right\}\ ,
\eqno(13)
$$
in which 
$$
\eqalignno{
I_\omega(\Lambda;L)&=\int d^{2\omega-1}{\bf k}\int_{-\infty}^{+\infty} dk_3\
{\sin^2(k_3L)\over {\bf k}^2+\Lambda^2}\left(
{1\over k_3^2}-{1\over k_3^2+{\bf k}^2+\Lambda^2}\right)\ ,&(14a)\cr 
\hat I_\omega(\Lambda;T)&=\int d^{2\omega-1}{\bf k}\int_{-\infty}^{+\infty}
dk_4\ {\sin^2(k_4T)\over {\bf k}^2}\left(
{1\over k_4^2+\Lambda^2}-{1\over k_4^2+{\bf k}^2+\Lambda^2}\right)\ ,&(14b)\cr 
J_\omega(\Lambda;L,T)&=\int_{-\infty}^{+\infty} dk_3\int_{-\infty}^{+\infty}
dk_4 \int d^{2\omega-2}k_\perp\ {\sin^2(k_3L)\over k_3^2}
{\cos(2k_4T)\over k_\perp^2+k_3^2+k_4^2+\Lambda^2}\ ,&(14c)\cr
\hat J_\omega(\Lambda;T,L)&=\int_{-\infty}^{+\infty}
dk_3\int_{-\infty}^{+\infty} dk_4 \int d^{2\omega-2}k_\perp\
{\sin^2(k_4T)\over k_4^2+\Lambda^2} {\cos(2k_3L)\over
k_\perp^2+k_3^2+k_4^2+\Lambda^2}\ .&(14d)\cr} 
$$
Explicit calculation yields [10]
$$
\eqalignno{
I_\omega(\Lambda;L)&=
(L\Lambda)\Lambda^{2\omega-4}\pi^\omega\sqrt{\pi}\ \Gamma\left({3\over
2}-\omega\right)-\Lambda^{2\omega-4}\pi^\omega\Gamma(2-\omega)\cr
&+\Lambda^{2\omega-4}{2\pi^\omega\over 2\omega-3} (\Lambda L)^{2-\omega}\ 
K_{2-\omega}(2\Lambda L)\cr
&-\Lambda^{2\omega-4}{\pi^\omega\sqrt{\pi}\over 2\Gamma\left(\omega-{1\over
2}\right)} \int_0^\infty{dx\over \sqrt{x}}\ {x^{\omega-2}\over
(1+x)^{3/2}}\exp\{-2\Lambda L\sqrt{1+x}\}\ ,&(15a)\cr
\hat I_\omega(\Lambda;T)&=\Lambda^{2\omega-4}{\pi^\omega\over 2\omega-3}
\left\{2(\Lambda T)^{2-\omega}K_{2-\omega}(2\Lambda T)-
\Gamma(2-\omega)\right\}\ ,&(15b)\cr
J_\omega(\Lambda;L,T)&={\pi^\omega\over \sqrt{\pi T\Lambda}}\left({T\over
\Lambda}\right)^{2-\omega}\sum_{n=1}^\infty (-)^{n+1}{(2\Lambda L)^{2n}\over
(2n)!}\cr
&\times\int_0^\infty dx\
x^{n-3/2}\left(\sqrt{1+x}\right)^{\omega-3/2}K_{\omega -3/2}(2\Lambda
T\sqrt{1+x})\ ,&(15c)\cr 
\hat J_\omega(\Lambda;L,T)&={\pi^\omega\over \sqrt{\pi L\Lambda}}\left({L\over
\Lambda}\right)^{2-\omega}\sum_{n=1}^\infty (-)^{n+1}{(2\Lambda T)^{2n}\over
(2n)!}\cr
&\times\int_0^\infty dx\
x^{n-1/2}\left(\sqrt{1+x}\right)^{\omega-7/2}K_{\omega -3/2}(2\Lambda
L\sqrt{1+x})\ ,&(15d)\cr}
$$
where $K_\nu(z)$ denotes the Bessel function of imaginary argument [10].
The ${\cal O}(g^2)$ result around $\omega=2$ in this case
can be rewritten as
$$
{\tt reg}\hat W_2(\Lambda;L,T)\buildrel \omega\uparrow 2 \over \sim
W_2^{\rm div}+\hat W_2^{\rm fin}(\Lambda;L,T)\ ,
\eqno(16)
$$
$$
\eqalign{
\hat W_2^{\rm fin}(\Lambda;L,T)
& =2{\hat g}^2 C_2(F)\left\{\pi\Lambda L
-\gamma_E-\ln\left({\Lambda^2\over 4\pi\mu^2}\right)\right.\cr
& -K_0(2\Lambda L)-K_0(2\Lambda T)+{1\over 2}{\cal I}_0(2\Lambda L)+
{1\over 2}\cr
&+\left.\sum_{n=1}^\infty{(-)^{n+1}\over 2(2n)!}\left[
(2\Lambda L)^{2n}{\cal I}_n(2\Lambda T)+
(2\Lambda T)^{2n}\hat{\cal I}_n(2\Lambda L)
\right]\right\}\ ,\cr}
\eqno(17)
$$
where we have set
$$
\eqalignno{
&{\cal I}_0(2\Lambda L)\equiv \int_0^\infty{dx\over \sqrt{x}}\ 
(1+x)^{-3/2}\exp\{-2\Lambda L\sqrt{1+x}\}\ ,&(18a)\cr
&{\cal I}_n(2\Lambda L)\equiv {1\over 2\Lambda L}
\int_0^\infty dx\ x^{n-3/2}\exp\{-2\Lambda L\sqrt{1+x}\}\ ,\qquad
n\in{\bf N}\ ,&(18b)\cr
&\hat{\cal I}_n(2\Lambda L)\equiv {1\over 2\Lambda L}
\int_0^\infty dx\ {x^{n-1/2}\over 1+x}\exp\{-2\Lambda L\sqrt{1+x}\}\ ,\qquad
n\in{\bf N}\ .&(18c)}
$$
Again, after summing up the mass independent counterterm
$$
\hat W_2^{\rm c.t.}=-2\hat g^2C_2(F)\left({1\over
\epsilon}+\gamma_E+{3\over 2}+\ln\pi\right)\ , 
\eqno(19)
$$
we finally get the minimal form of the ${\cal O}(g^2)$ ultraviolet renormalized and 
infrared cut off  Wilson loop in the axial gauge: namely,
$$
\eqalign{
\hat W_2^{\rm ren}(\Lambda;L,T)
& =2{\hat g}^2 C_2(F)\left\{\pi\Lambda L
-1-\ln\left({\Lambda^2\over 4\mu^2}\right)\right.\cr
& -K_0(2\Lambda L)-K_0(2\Lambda T)+{1\over 2}{\cal I}_0(2\Lambda L)\cr
&+\left.\sum_{n=1}^\infty{(-)^{n+1}\over 2(2n)!}\left[
(2\Lambda L)^{2n}{\cal I}_n(2\Lambda T)+
(2\Lambda T)^{2n}\hat{\cal I}_n(2\Lambda L)
\right]\right\}\ ,\cr}
\eqno(20)
$$
in such a way that the ${\cal
O}(g^2)$ renormalized quantities in their minimal forms fulfill
$$
%\lim_{\Lambda\to 0}W_2^{\rm ren}(\Lambda;L,T)=d
\lim_{\Lambda\to 0}\hat W_2^{\rm ren}(\Lambda;L,T)
=W_2^{\rm ren}(\beta,\sigma)\ .
\eqno(21)
$$
The content of the lowest order result (21) is what could be reasonably expected: 
the renormalized infrared cut off Wilson loop in the axial gauge
flows smoothly to the corresponding gauge invariant value, once the infrared cut off 
is removed and up to renormalization prescriptions. 
It is important to stress that the limits $\omega\uparrow 2$ and $\Lambda\downarrow 0$ do not commute and,
consequently, in order to obtain the finite and consistent result (21) one has to 
first subtract the ultraviolet infinities, then take the limit
$\omega\uparrow 2$ and finally the limit $\Lambda\downarrow 0$. 
A further observation is that by performing the very same
calculation for the massive case in a general linear covariant gauge, 
the same result (21) is obtained once again, up to some slightly different 
renormalization prescription. 
\bigskip
\noindent
{\bf 4.}\quad Let us now turn our attention to the highly non-trivial ${\cal O}(g^4)$ calculation.
It is known [11] that the verification of the large $T$-limit exponential behaviour of the euclidean 
Wilson loop in perturbation theory 
does already provide a crucial consistency test of temporal and axial gauges quantization schemes 
for non-abelian massless gauge theories.
In the previous section
we have shown that, at the lowest order ${\cal O}(g^2)$, the 
renormalization group flow of the infrared cut off Wilson loop in the axial gauge converges smoothly, 
after subtraction of ultraviolet infinities,
to the gauge invariant result, in the limit of the removal of the infrared cut off $\Lambda$ - 
see eq.~(21). In the present section we want to check whether
the very same conclusion is true at the next order ${\cal O}(g^4)$.
If this were true, then the RFG approach to the
quantization of non-abelian gauge theories, such as proposed in ref.s~[5,6], 
would be extremely appealing. Otherwise, it is
ruled out. The gauge invariance test consists of two parts: first, one has to 
check that all the infrared singularities do cancel
\footnote{$^1$}{We stress once again that bad infrared divergences do definitely appear 
in individual graphs, as already manifest in the 
${\cal O}(g^2)$ calculation, because the free propagator (2)
does not admit a limit $\Lambda\downarrow 0$ in the tempered distributions weak topology.}
and second, in such a case, 
one has to verify that the ultraviolet and infrared finite part does coincide, 
up to renormalization prescriptions,  
with the corresponding expression as computed, {\it e.g.}, in the Feynman's gauge.

In order to extract the infrared singular part of any individual graph, the following 
{\it Lemma} is fairly useful.
We start from the identity
$$
{1\over x^2+\Lambda^2}= -{d\over dx}\left({x\over x^2+\Lambda^2}\right)+
{2\Lambda^2\over (x^2+\Lambda^2)^2}\ ,
\eqno(22)
$$
whence we obtain that, for any continuous test function $F(x;\Lambda),\ x\in{\bf
R},\ \Lambda\ge 0$, the following relationship holds true: namely,
$$
\eqalign{
\int_{-\infty}^{+\infty}dx\ {F(x;\Lambda)\over x^2+\Lambda^2}&=
\int_{-\infty}^{+\infty}dx\ F(x;\Lambda)\left(-{d\over dx}{x\over
x^2+\Lambda^2}\right)\cr
&+{1\over \Lambda}\int_{-\infty}^{+\infty}dx\ F(x;0)
{2\Lambda^3\over (x^2+\Lambda^2)^2}\cr
&+\int_{-\infty}^{+\infty}dx\ [F(x;\Lambda)-F(x;0)]
{2\Lambda^2\over (x^2+\Lambda^2)^2}\ .\cr}
\eqno(23)
$$
Now, according to the theory of
distributions [12] we have
$$
\eqalignno{
&{\cal S}^\prime\lim_{\Lambda\downarrow 0}
{x\over x^2+\Lambda^2}={\rm CPV}\left({1\over x}\right)
\equiv {1\over [x]}\ ,&(24a)\cr
&{\cal S}^\prime\lim_{\Lambda\downarrow 0}
\left(-{d\over dx}\right){x\over x^2+\Lambda^2}=
\left(-{d\over dx}\right){\rm CPV}\left({1\over
x}\right) \equiv {1\over \left[x^2\right]}\ ,&(24b)\cr
&{\cal S}^\prime\lim_{\Lambda\downarrow 0}
{2\Lambda^3\over \pi (x^2+\Lambda^2)^2}=
{\cal S}^\prime\lim_{\Lambda\downarrow 0}
{\Lambda\over \pi (x^2+\Lambda^2)}=\delta(x)\ ,&(24c)
\cr}
$$
where ${\cal S}^\prime\lim$ means limit in the tempered distrubutions topology and CPV indicates the Cauchy Principal Value.
Therefrom, under the assumption that
$$
F(x;\Lambda)-F(x;0)=\Lambda^2 f(x;\Lambda)\ ,\qquad x\in{\bf R}\ ,
\eqno(25)
$$
with $f(x;\Lambda)$ analytic when $\Lambda\downarrow 0$,
one finds the following small $\Lambda$
behaviour 
$$
\int_{-\infty}^{+\infty}dx\ {F(x;\Lambda)\over x^2+\Lambda^2}\ \buildrel 
\Lambda\downarrow 0 \over \sim\ 
{\pi\over \Lambda} F(0;0)
%+\pi\hat F^\prime_\Lambda(0)
+\int_{-\infty}^{+\infty}dx\ {F(x;0)\over \left[x^2\right]}
+{\cal O}(\Lambda)\ .
\eqno(26)
$$
The above formula (26) is the basic tool to single out the infrared divergences 
of the different graphs contributing to the
${\cal O}(g^4)$ expression of the infrared cut off Wilson loop in the axial gauge. 
The whole set of diagrams contributing to the 
genuine non-abelian part of the dimensionally regularized infrared cut off Wilson loop in the axial gauge -
which are proportional to the product of the quadratic Casimir operators
$C_2(A)C_2(F)$ of the adjoint and fundamental representations of $SU(N)$ -  
can be suitably grouped into three classes we shall call: 
$(a)$ the self-energy diagrams (fig.1); $(b)$ the crossed propagators diagrams 
(fig.2); $(c)$ the three point diagrams (fig.3).
Let us start by checking whether the strongest\footnote{$^2$}{Actually, 
it is not difficult to verify that all the terms 
potentially leading to ${\cal O}(\Lambda^{-3})$ infrared singularities do identically vanish.}
infrared singularity ${\cal O}(\Lambda^{-2})$ of the non-abelian part of the infrared cut off Wilson loop
does cancel or not in the limit $\Lambda\downarrow 0$. 
\medskip\noindent
$(a)$\quad The contribution of the self-energy diagrams reads
$$
\eqalign{
{\tt reg}{\overline W}_4^{{\rm (a)}}(\Lambda;T,L)&=
-{\hat g^4\over 8\pi^4}C_2(A)C_2(F)\int_p\int_k{\sin^2(p_3L)\over
p_3^2}{1-\cos(2p_4T)\over (p^2+\Lambda^2)^2}\cr
&\times\left(\delta_{i3}+{p_ip_3\over p_4^2+\Lambda^2}\right)
\left(\delta_{j3}+{p_jp_3\over p_4^2+\Lambda^2}\right)
{1\over (k^2+\Lambda^2)[(p-k)^2+\Lambda^2]}\cr
&\times\left[\delta_{mn}+{k_mk_n\over k_4^2+\Lambda^2}\right]
\left[\delta_{lr}+{(p-k)_l(p-k)_r\over (p_4-k_4)^2+\Lambda^2}\right]
V_{irm}V_{jln}\ ,\cr}
\eqno(27)
$$
where latin indices take values 1,2,3 (remember that the gauge vector is here supposed to have components
$n_i=0,n_4=1$) and we have set for later convenience
$$
\int_p\equiv \int{d^{2\omega}p\over
(2\pi\mu)^{2\omega-4}}\ ,
$$
whereas
$$
 V_{irm}\equiv (p-2k)_i\delta_{mr}+(k-2p)_m\delta_{ir}+(p+k)_r\delta_{im}\ .
\eqno(28)
$$
It can be readily checked that
$$
p_i(p-k)_rk_mV_{irm}=0\ ,\qquad p_j(p-k)_lk_nV_{jln}=0\ ,
\eqno(29)
$$ 
whence it follows that the infrared most singular part of the self-energy diagrams turns out to be
$$ 
\left.{\tt reg}{\overline W}_4^{{\rm (a)}}(\Lambda;T,L)\right|_{\rm sing}
=-{\hat g^4\over 2\pi^4}C_2(A)C_2(F)\left[
{\cal I}_1^{\rm (a)}(\Lambda;L,T)+{\cal I}_2^{\rm (a)}(\Lambda;L,T)\right]\ ,
\eqno(30)
$$
where
$$
\eqalignno{
{\cal I}_1^{\rm (a)}(\Lambda;L,T)&\equiv
\int_p
{\sin^2(p_4T)\over
(p_4^2+\Lambda^2)^2}{\sin^2(p_3L)\over (p^2+\Lambda^2)^2}\cr
&\times\int_k
{({\bf p}-{\bf k})^2[{\bf p}^2{\bf k}^2-({\bf p}\cdot{\bf k})^2]\over
(k_4^2+\Lambda^2)(k^2+\Lambda^2)[(p-k)^2+\Lambda^2]}\ ,&(31a)\cr
{\cal I}_2^{\rm (a)}(\Lambda;L,T)&\equiv
\int_p
{\sin^2(p_4T)\over
(p^2+\Lambda^2)^2}{\sin^2(p_3L)\over 2p_3^2}\cr
&\times\int_k
{({\bf p}^2)^2 k_3^2+({\bf p}\cdot{\bf k})^2 p_3^2 -2p_3 k_3({\bf p}\cdot{\bf
k}){\bf p}^2\over
(k^2+\Lambda^2)[(p-k)^2+\Lambda^2](k_4^2+\Lambda^2)[(p_4-k_4)^2+\Lambda^2]}\ .&(31b)\cr}
$$
A direct inspection of the integrals (31) shows, taking the previously quoted 
{\it Lemma} carefully into account,
that the self-energy diagrams do not produce any ${\cal O}(\Lambda^{-2})$ 
infrared singularities, or equivalently
$$
\lim_{\Lambda\downarrow 0}(\Lambda^2){\tt reg}{\overline W}_4^{{\rm (a)}}(\Lambda;T,L)=0\ ,
\eqno(32)
$$
as the sum of the self-energy diagrams contains at most ${\cal O}(\Lambda^{-1})$ infrared singularities. 
\medskip\noindent
$(b)$\quad The contribution of the crossed propagators diagrams reads
$$
\eqalign{
{\tt reg}{\overline W}_4^{{\rm (b)}}(\Lambda;T,L)&={\hat g^4\over 16\pi^4}C_2(A)C_2(F)\int_p\int_q
\tilde D_{33}(p;\Lambda,n)\tilde D_{33}(q;\Lambda,n)\cr
&\times\left\{ 2\exp\{-2iT(p_4+q_4)\}\left[{\sin[(p_3+q_3) L]\over p_3 (p_3+q_3)} -
\exp\{-ip_3 L\}{\sin(q_3 L)\over p_3 q_3}\right]^2\right.\cr
&-2\exp\{-2iq_4 T\}{\sin(q_3 L)\over q_3}{\cal F}(p_3,q_3;L)\cr
&+2\exp\{2iq_4 T\}{\sin(q_3 L)\over q_3}{\cal F}(p_3,q_3; -L)\cr
&+\left.{\sin^2[(p_3+q_3)L]\over p_3 q_3(p_3+q_3)^2}+{q_3\sin^2(p_3 L)
+p_3\sin^2(q_3 L)\over p_3^2 q_3^2 (p_3+q_3)}\right\}\ ,\cr}
\eqno(33)
$$
where
$$
{\cal F}(p_3,q_3;L)\equiv {\sin(q_3L)\over p_3q_3(p_3+q_3)}
-{\sin[(p_3+q_3)L]\over p_3q_3(p_3-q_3)}
+\exp\{-i(p_3+q_3)L\}{\sin(p_3L)\over p_3q_3(p_3+q_3)}\ .
\eqno(34)
$$
A straightforward evaluation, taking again the {\it Lemma} into account, actually shows that the expression in eq.~(33)
does not produce any ${\cal O}(\Lambda^{-2})$ singularity: namely,
$$
\lim_{\Lambda\downarrow 0}(\Lambda^2){\tt reg}{\overline W}_4^{{\rm (b)}}(\Lambda;T,L)=0\ .
\eqno(35)
$$
\medskip\noindent
$(c)$\quad The contribution of the three point diagrams reads
$$
\eqalign{
{\tt reg}{\overline W}_4^{{\rm (c)}}(\Lambda;T,L)&=
{4i(ig)^3\over N(2\pi)^8}{\tt tr}(T^A_F T^B_F T^C_F)
\int_p\int_q\int_r\tilde G_{333}^{ABC}(p,q,r)\delta^{(2\omega)}(p+q+r)
\cr
&\times (2\pi\mu)^{2\omega-4}\left\{2\cos(2p_4 T){\sin^2(p_3 L)\over p_3^2 r_3}+
{\sin^2(p_3 L)\over p_3^2 q_3} -{\sin^2(r_3 L)\over r_3^2 q_3}\right\}\ ,\cr}
\eqno(36)
$$
where $\tilde G_{\mu\nu\rho}^{ABC}(p,q,r)$ denotes the three-point Schwinger's function in momentum space.
A straightforward calculation eventually drives to the identification of 
the infrared most singular part of the expression in eq.~(36): namely,
$$
\eqalign{
\left.{\tt reg}{\overline W}_4^{{\rm (c)}}(\Lambda;T,L)\right|_{\rm sing}&= 
{\hat g^4\over (4\pi\Lambda)^2}C_2(A) C_2(F)\int_{\bf q}\int_{\bf r}({\bf q}^2{\bf r}^2)^{-1}\cr
&\times\left\{\sin^2[(q_3+r_3)L]{(q_3-r_3)^2\over (q_3+r_3)^2}
-4{q_3\sin^2(r_3 L)\over q_3-r_3}\right\}\ ,\cr}
\eqno(37)
$$
with 
$$
\int_{\bf p}\equiv \int d^{2\omega-1}{\bf p}\
(2\pi\mu)^{4-2\omega}\ .
$$ 
From the values of the basic dimensionally regularized integrals
$$
\eqalignno{
\lim_{\omega\uparrow 2}\ &\int_{\bf p}{\sin^2(p_3 L)\over {\bf p}^2}=-{\pi^2\over 2L}\ ,&(38a)\cr
\lim_{\omega\uparrow 2}\ &\int_{\bf p}\int_{\bf q}{p_3 q_3\sin^2[(p_3+q_3)L]\over
{\bf p}^2{\bf q}^2(p_3+q_3)^2}=-{\pi^4\over 12L^2}\ ,&(38b)\cr
\Phi_\omega(\mu L)&\equiv 
 {L^2\over \pi^4}\int_{\bf p}\int_{\bf q}{(p_3+q_3) \sin[(p_3+q_3)L] \sin[(p_3-q_3)L]\over
{\bf p}^2{\bf q}^2(p_3-q_3)}\cr
&=\Gamma(2-\omega){(\pi/2)^{2\omega-4}(2\pi\mu L)^{8-4\omega}\Gamma\left({5\over 2}-\omega\right)\over
(3-2\omega)[\Gamma(\omega-1)]^2\Gamma\left({7\over 2}-2\omega\right)}\ ,&(38c)\cr}
$$
it follows that
$$
{\tt reg}{\overline W}_4^{{\rm (c)}}(\Lambda;T,L)
\ \buildrel 
\Lambda\downarrow 0 \over \sim\ 
{g^4\over (16\pi\Lambda L)^2} C_2(A) C_2(F)
\left\{\Phi_\omega(\mu L)-{1\over 6}\right\}\ .
\eqno(39)
$$
Summing up the results of eq.s~(32),(35),(39) we definitely find the following 
behaviour of the non-abelian part of the 
${\cal O}(g^4)$ dimensionally regularized and infrared cut off Wilson loop in the axial gauge, {\it i.e.},
$$
\eqalign{
{\tt reg}{\overline W}_4(\Lambda;T,L)&\equiv
{\tt reg}{\overline W}_4^{{\rm (a)}}(\Lambda;T,L)+
{\tt reg}{\overline W}_4^{{\rm (b)}}(\Lambda;T,L)+
{\tt reg}{\overline W}_4^{{\rm (c)}}(\Lambda;T,L)\cr
&\buildrel 
\Lambda\downarrow 0 \over \sim\ {g^4\over (16\pi\Lambda L)^2} C_2(A) C_2(F)
\left\{\Phi_\omega(\mu L)-{1\over 6}\right\}\ .\cr}
\eqno(40)
$$
\bigskip\noindent
{\bf 5.}\quad
The result of eq.~(40) shows that the ${\cal O}(g^4)$ dimensionally regularized and 
infrared cut off Wilson loop in the axial gauge
does not converge to some infrared  finite result at the physical point $\Lambda=0$. 
This means, in turn, that 
the Renormalization Group Flow in the axial gauge is discontinuous at the physical point, even for a formally 
gauge invariant quantity such as the euclidean Wilson loop (this conclusion is {\it a fortiori} manifestly also true
for the Schwinger's functions). Eq.~(40) also shows that the gauge invariance breaking term is extremely bad, since it
corresponds to the product of infrared and ultraviolet singularities, the RHS of eq.~(40) being
${\cal O}\left(\Lambda^{-2}(2-\omega)^{-1}\right)$. 

The example we have worked out in this note
clearly indicates that the RGF approach to the non-abelian Yang-Mills theories in the axial gauge
is absolutely unreliable and, in turn, the proposals suggested in ref.s~[5,7] are definitely ruled out at least
within perturbation theory. To this concern, it is worthwhile to stress once again that the RGF approach to Yang-Mills theories
in Minkowski space-time and 
in the axial gauge manifestly fulfills, for any $\Lambda^2>0$, power counting renormalizability, 
unitarity - there are only two field polarizations with positive definite metric - and minimal breaking of
gauge invariance - very simple modified Lee-Ward-Takahashi identities hold true to all scales.
The whole set of those basic requirements, which make perturbation theory perfectly well defined to all orders,
is achieved at the price of breaking simultaneously both gauge invariance - owing to the presence of the infrared cut
off $\Lambda$ - and Lorentz invariance - owing to the presence of the gauge four-vector $n_\mu$.
The meaning of the calculation we have developed in the present note  is that 
the RGF approach to non-abelian gauge theories in the axial gauge is physically inconsistent in perturbation theory,
as gauge and Lorentz invariances can not be generally recovered in the physical limit, owing to the presence of
bad infrared singularities. 
The wild singularity at the physical point 
of a formally gauge invariant quantity in the axial gauge also suggests, in our opinion, that 
the assumption of the regularity at the physical point $\Lambda\downarrow 0$
of the RGF for observable physical quantities in a non-abelian gauge theory
is a highly non-trivial requirement, which should be better investigated and understood, in general, 
also within covariant and light-cone gauge choices [7]. 
\bigskip\noindent
{\bf Acknowledgements}
\medskip\noindent
We thank C. Becchi, G.C. Rossi, M. Simionato and M. Testa for quite useful discussions. 
We also thank P. Giacconi for her help during the early stage of the calculations. 
\bigskip
\noindent
{\bf References}
\medskip
\item{[1]} K. G. Wilson, Phys. Rev. B{\bf 4} (1971) 3174;
K. G. Wilson and I. G. Kogut, Phys. Rep. {\bf 12} (1974) 75.
\item{[2]} J. Polchinski, Nucl. Phys. B{\bf 231} (1984) 269.
\item{[3]} For a quite complete and up to date list of references see {\it e.g.}:
C. Bagnuls and C. Bervillier, {\tt hep-th/0002034}.
\item{[4]} C. Becchi, in {\it Elementary Particles, Field Theory and
Statistical Mechanics}, M. Bonini, G. Marchesini and E. Onofri Eds., Parma
University 1993,  {\tt hep-th/9607188}; M. Bonini, M. D'Attanasio and G. Marchesini, Nucl. Phys. B{\bf 437} (1995) 163;
{\it ibid.} B{\bf 444} (1995) 602.
\item{[5]} D. F. Litim and J. M. Pawlosky, Phys. Lett. B{\bf 435} (1998) 181;
Nucl. Phys. Proc. Suppl. {\bf 74} (1998) 329.
\item{[6]} M. Simionato, Int. J. Mod. Phys. A{\bf 15} (2000) 2121, {\it ibid.} 2153.
\item{[7]} M. Simionato, {\tt hep-th/0005083} 
\item{[8]} A. Bassetto, G. Nardelli and R. Soldati, {\it Yang Mills Theories
in algebraic non-covariant gauges}, World Scientific, Singapore (1991) p. 80-85.
\item{[9]} See {\it e.g.} the textbook: M. E. Peskin and D. V. Schroeder, {\it
An Introduction to Quantum Field Theory}, Perseus, Reading (1995) p. 503.
\item{[10]} I. S. Gradshteyn and I. M. Ryzhik, {\it Table of Integrals, Series and Products}, Academic Press,
San Diego (1980).
\item{[11]} G. C. Rossi and M. Testa, Nucl. Phys. B{\bf 163} (1980) 109; {\it ibid.} B{\bf 176} (1980) 477;
S. Caracciolo, G. Curci and P. Menotti, Phys. Lett. B{\bf 113} (1982) 311;
J. P. Leroy, J. Micheli and G. C. Rossi, Nucl. Phys. B{\bf 232} (1984) 511;
P. V. Landshoff, Phys. Lett. B{\bf 169} (1986) 69; 
H. H\"uffel, P. V. Landshoff and J. C. Taylor, Phys. Lett. B{\bf 217} (1989) 147;
G. Nardelli and R. Soldati, Int. J. Mod. Phys. {\bf 5} (1990) 3171.
\item{[12]} I. M. Guelfand and G. E. Chilov, {\it Les Distributions - Tome I}, Dunod, Paris (1962).
\vfill\eject\end